\documentclass[letterpaper, 10 pt, conference]{ieeeconf}  

\IEEEoverridecommandlockouts                              

\usepackage{graphics} 
\usepackage{epsfig} 
\usepackage{mathptmx} 
\usepackage{times} 
\usepackage{amsmath} 
\usepackage{amssymb}  
\usepackage{gensymb}
\usepackage{multirow}
\usepackage{tabularx}
\usepackage{makecell}
 
\usepackage{dirtytalk}
\usepackage[normalem]{ulem}
\usepackage{xcolor}
\usepackage{wrapfig}

\usepackage{graphicx}
\DeclareMathOperator{\atantwo}{atan2}
\usepackage{float}
\usepackage{lipsum}
\usepackage{algorithm}
\usepackage{algorithmic}
\usepackage[font=small,labelfont=bf, skip=2pt]{caption}
\usepackage{subcaption}
\usepackage{etoolbox}
\usepackage{hyperref}
\hypersetup{
    colorlinks,
    citecolor=black,
    filecolor=black,
    linkcolor=blue,
    urlcolor=black
}

\usepackage{titlesec}
\titlespacing*{\section}
{0pt}{1.5mm}{1mm}
\titlespacing*{\subsection}
{0pt}{0.9mm}{0.9mm}

\usepackage{multicol}
\usepackage[british]{babel}
\usepackage[autostyle]{csquotes}

\title{\LARGE \bf
Control Interface Remapping for Bias-Aware Assistive Teleoperation
}

\begin{document}

\author{Andrew Thompson$^{*^{1,3}}$, Larisa Y.C. Loke$^{*^{1,3}}$, Brenna Argall$^{1,2,3}$
\thanks{*Authors contributed equally to this work}
\thanks{$^{1}$Department of Mechanical Engineering, Northwestern University, Evanston, IL, USA}%
\thanks{$^{2}$Departments of Computer Science and Physical Medicine \& Rehabilitation, Northwestern University, Evanston, IL, USA}%
\thanks{$^{3}$Shirley Ryan AbilityLab, Chicago, IL, USA}
}

\maketitle
\thispagestyle{empty}
\pagestyle{empty}

\begin{abstract}
Users of assistive devices vary in their extent of motor impairment, and hence their physical interaction with control interfaces can differ. There is the potential for improved utility if
control interface actuation is mapped to assistive device control signals in a manner customized to each user. 
In this paper, we present (1) a method for creating a custom interface to assistive device control mapping based on the design of a user's bias profile, (2) a procedure and virtual task for gathering interface actuation data from which to build the bias profile and map, and (3) an evaluation of our method on 6 participants with upper limb motor impairments.
Our results show that custom interface remapping based on user bias profiles shows promise in providing assistance via an improvement in the reachability of the device control space. This effect was especially pronounced for individuals who had a more limited reachable space. 

\end{abstract}

\section{INTRODUCTION}

There is a large variability in users' interactions with control interfaces used to engage with robotic and assistive 
systems. Despite efforts to make this interaction more seamless~\cite{Abbink2018}, there persists a lack of truly individualized 
customization of most interfaces. 
Within the domain of assistive robotics, this discrepancy is especially pronounced due to varying degrees of injury and recovery among the intended users. 
When it limits the controllability afforded to users with motor impairment, the result can be
a lack of user agency
that results in frustration, disuse, and even injury~\cite{Parasuraman1997a}. 

The question of modeling \textit{how} an individual uses an interface is an important one, as the information encoded in such a model can be used to provide customized assistance in accordance with personalized interface usage characteristics and preferences. 
In this work, we address bias in teleoperation that stems from a user's ability (or inability) to reach the entirety of an interface's control space, and which is influenced by a given user's behavioral tendencies. Computing such a bias metric allows for comparison to an ideal interface actuation baseline, 
defined to be the entire available control space respective to the chosen interface. 

The ability to compare a specific user's bias against this ideal baseline is beneficial for a number of reasons: (1) specific to this study, this comparison allows us to remap the regions of the control space that the user is able to reach to then encompass the entire available control space; (2) knowledge of asymmetric patterns in teleoperation may be used to help influence training in the use of that interface, tailored to the individual user; (3) this information could potentially be utilized as an additional evaluation technique for approving individuals to operate various assistive devices; and (4) the computation of asymmetric divergence between baseline and user sets situates our remapping technique for insertion into a machine learning framework, should the divergence be used as a term to guide optimization.

Since individuals often experience day-to-day variation 
in their behavior due to any number of internal and environmental factors, and because individuals who operate assistive devices often utilize their devices in multiple settings, it makes sense that a single mapping for an interface might not be comprehensive for all scenarios. To address this, we present our mapping procedure in a framework called \textit{Augmented Dynamic Remapping} (ADR), which allows for context-dependent swapping between control maps in real-time, with a tunable degree of transitional smoothing. ADR could allow, for example, a user operating an assistive device to seamlessly transition from an \textquote{at-home} mapping to a \textquote{street} mapping; alternatively, ADR could be used to further segment an \textquote{at-home} mapping into task- or room-specific maps that arise in response to the user's daily activities.

This work presents the following contributions:
\begin{enumerate}
    \item \textit{Design of a user's bias profile:} An individual user's bias profile is a way to quantitatively compare their task-agnostic skill in teleoperation to
    other users and against baseline or expert conditions. We utilize well-established geometric optimization techniques to characterize user bias during teleoperation.
    \item \textit{Augmented Dynamic Remapping (ADR):} ADR is a method which builds a customized map between a user's control interface actuation and the space of allowable control signals. It allows for a real-time swap between different remappings of the control space according to context information during teleoperation; we use this in conjunction with various bias profiles to allow for user-specific remapping. 
    \item \textit{Mitigation of signal noise via bias models:}
    With a model of a user's bias across multiple tasks, we use this customized model of the user in order to
    recover approximate control signals in instances of control signal drop-out or the presence of heavy noise.
\end{enumerate}

In Section~\ref{sec:background}, we discuss some related works and general background for the challenges outlined above. In Sections~\ref{sec:bias_metric_formulation}--\ref{sec:bias_metric_deployment}, we define our bias metric and provide insight on our specific approach. In Section~\ref{sec:exp_design}, we detail the structure of our experiment. In Sections~\ref{sec:results_and_discussion}--\ref{sec:discussion}, we present the data from our study and offer some results and explanatory discussion. 

\section{BACKGROUND}\label{sec:background} 

One difficulty in designing assistive devices for individuals with motor impairment is that the design objective is often in flux. That is---due to the effects of degeneration and rehabilitation---the extent to which an individual is affected by their motor impairment on any given day is bound to shift
\cite{engelberg1996musculoskeletal}. Depending on the severity of these shifts, users might experience variation over time in their ability to reach the entire control space. Even a transient inability to reach the entire control space can limit the options of control interfaces available for individuals with neuromotor deficits; and any sort of reduction in control space reachability
inherently limits the commands that can be issued with that interface. 

Longstanding techniques for customizing
assistive device control interfaces 
adjust certain preset mechanical configurations and thresholds. These techniques, outlined in~Guirand et al.~\cite{guirand2011tuning}, 
include uniform scaling along principal axes, axial tilt, and radial expansion/contraction of deadzones. They
often require expert clinicians to complete, and/or additional measurement equipment or hardware to accomplish~\cite{gillham2017feature}. 

In the domain of workspace spanning, similar approaches as utilized in our methodology have been proposed to map the effective workspace of two separate robots to one another~\cite{conti2005spanning}. It is less common for these approaches to be used at the level of the assistive interface, with one notable exception 
being when the interface and robot are one and the same~\cite{rossa2017nonlinear}.

There is a dissonance that occurs when the user of an assistive device does not feel as though they are the one in control of the device. A lack of a sense of agency when controlling an assistive device can lead to dissatisfaction, injury, or total disuse~\cite{Parasuraman1997a}. Such an effect also presents when control signals for the device derive from robotics autonomy paradigms that infer human intent~\cite{Abbink2018}, especially when human-autonomy transparency~\cite{lyons2013being} is lacking or incomplete.

To the authors' knowledge, there are currently no computationally-efficient procedures for designing a one-shot remapping of a given control interface (1) in an asymmetric and user-specific manner that preserves user agency, and (2) without the help of a skilled technician.

\section{BIAS METRIC FORMULATION}\label{sec:bias_metric_formulation}
Here we provide motivation, and a tractable representation, for bias metrics as a characterization of interface use.

\subsection{A Metric of Bias in Interface Use}

The variety of physiological user bias that we are interested in addressing with this work is a consequence of limited (in range or stability)
 interface actuation, 
 due to an individual's unique physiology. This bias presents as a systematic error between the control space accessible to the user and the full control space of the interface. Such bias
 is amplified to a greater degree in individuals with motor impairment. The reason, in part, is because interfaces generally are designed for a specific physiological average that does not address the variety of physiological states that may benefit from use of these devices. The concept of a quantifiable metric which conveys an individual's interface-dependent teleoperation bias is at the heart of the current work.

\begin{wrapfigure}[18]{R}{3.3cm}
    \centering
    \includegraphics[width=1\linewidth]{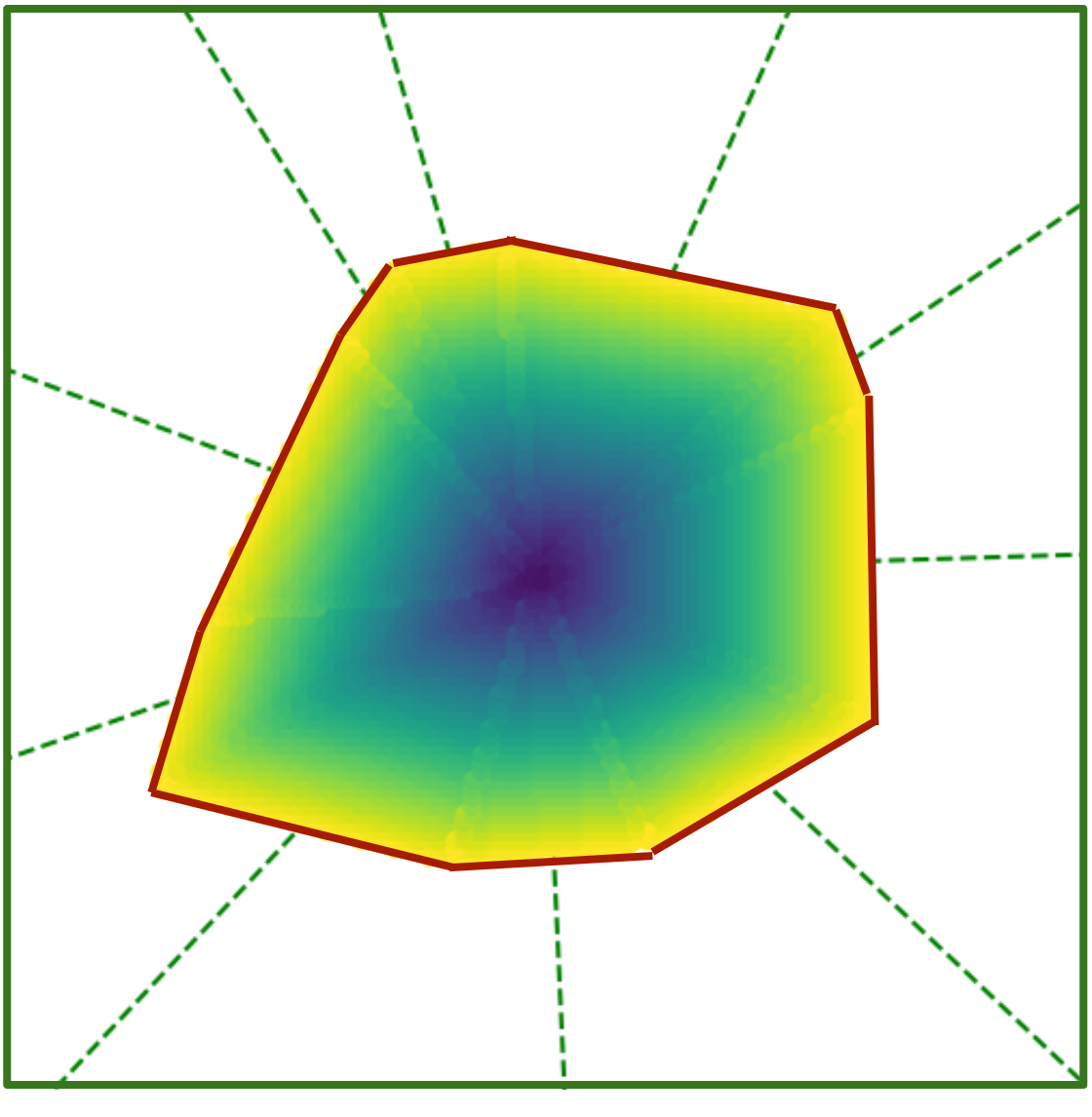}
    \caption{Example parameterization of 
    reachable control space using CHO. Convex hull (red outline) and its simplices (straight-line components). 
    Colors represent control value level sets within the hull.
    Ray-tracing (green dashed lines) accounts for asymmetrical control distributions, further detailed in Section~\ref{sec:Dynamic Interface Remapping}.}
    \label{fig:2d_remap}
\end{wrapfigure}

Consider a 3-axis joystick, the interface used in our experimental work: a lever arm designed to be mechanically deflected about a radial pivot point at its base and simultaneously twisted independent of mechanical deflections of the lever-arm.
When mapped to control
a machine,
the maximum amount of deflection of the joystick about these three axes defines the bounds of the control space, and consequently the bounds of commandable control signals (such as velocity).

For those with motor impairments, 
there may be regions of the control space that are not reachable. We define the reachable set $\mathbb{U}_{\text{reach}}\subset\mathbb{U}$ for an individual  as the subset of the entire  control space $\mathbb{U}$ 
that is reachable by them  when operating a given interface. This reachable set can be determined in a number of ways; ours is described 
in Section~\ref{sec:Meaning of Reachability}.
The embedded, reachable region  subsequently can be compared to the morphology of the given interface’s entire control space, from which we can derive what we refer to as an individual’s personalized \textit{bias metric}.

\subsection{The Meaning and Parameterization of Reachability}\label{sec:Meaning of Reachability}

We define \textit{reachability}
as the set of all points in the interface control space that a user is physically able to access in an \textit{intentional} manner. For the 3-axis joystick, this reachable control set is embedded in the bounded ($u_x$, $u_y$, $u_z$)-space of possible velocity commands. 

We \textit{measure} reachability via temporal sampling, resulting in a sparse representation of a given user's reachable set. We \textit{parameterize} the reachable set using summary statistics---mean, variance, and omission.
\textit{Mean} and \textit{variance} are standards for the parameterization of normal (or near-normal) distributions of sampled data. \textit{Omission} refers to a combination of deadzones and rejection of unstable points
located on the outer bounds of the reachable control set.\footnote{The omission of unstable outliers, via what is effectively a low-pass filter, helps to reduce  
the effects of spasticity. For many assistive control interfaces such inputs already are filtered. Spastic motions are by nature unplanned, and our definition of reachability asserts that the reachable set $\mathbb{U}_{\text{reach}}$ is physiologically accessible in an intentional manner.}

In detail, we use convex hull estimation as a tractable parameterization
of the reachable set for a given user. \textit{Convex hull optimization} (CHO) algorithms~\cite{graham1983finding} return a set of linear functions, simplices $\mathbb{C}$, that together encapsulate the (majority) of points in a given set within a convex polygonal hull (Figure~\ref{fig:2d_remap}). Each vertex in the subsequent hull can be assigned a weight dependent on the number of control signals in their proximity. Control distributions with highly irregular geometries can thus be considered while also taking into consideration concentration of control signals,
 making CHO of particular interest in the categorization of interface bias.  A center of mass can be calculated for the distribution, along with a mean and variance along all axes. 
Parameter tuning changes the amount of points allowable outside 
the resultant hull; effectively filtering the set $\mathbb{U}_{\text{reach}}$.

We define the parameterization of bias $\xi$ using 2-dimensional CHO as:
\vspace{-2.5mm}
\begin{equation}\label{eq:Bias Parameterization via Convex Hull Optimization}
    \xi_{\tau} \equiv \{\mu_{u_{x}}, \sigma_{u_{x}}, \mu_{u_{y}}, \sigma_{u_{y}}, \tau_{x}^{cm}, \tau_{y}^{cm}\}, \nonumber
    \vspace{-2mm}
\end{equation}
where $\tau$ is a set of trajectories which spans the full control space, $\mu$ and $\sigma$ are the means and standard deviations of the set of control points in $\tau$ w.r.t. both of the x- and y-axes. The center of mass, denoted by $(\tau_{x}^{cm}, \tau_{y}^{cm})$, captures location bias in the user's preferred velocity profile; for instance, it might represent a tendency towards asymmetric teleoperation. The distribution metadata ($\mu, \sigma$) confers stability information which can be used for assessing whether a point should be included in the reachable set or otherwise omitted. 

\section{BIAS METRIC DEPLOYMENT\label{sec:bias_metric_deployment}}

In this section, we detail our specific implementation of the bias metric informed remapping, and formalize a number of the supporting concepts used in the remapping procedure. 

\subsection{Augmented Dynamic Remapping (ADR)} \label{sec:Dynamic Interface Remapping}

Given a representation, via CHO, of a user's reachable set $\mathbb{U}_{\text{reach}}$,                   
we then perform a dynamic remapping of the interface control signals. Our \textit{Augmented Dynamic Remapping (ADR)} approach compares the resulting hull to the entire available control space $\mathbb{U}$ in order to specify an augmented control map $\phi : \mathbb{U}_{\text{reach}} \to \mathbb{U}$ that \textit{stretches} the reachable space of the user to the full control space $\mathbb{U}$.

In detail, we first employ a discrete mapping method, that partitions $\mathbb{U}$ into a $m_x \times m_y$ cell grid. The level sets of this grid correspond to level sets of control signals with the same magnitude (though different directions), defined as $\mathbb{S}_{\mathbb{U}}$. Similarly, we partition the convex hull describing $\mathbb{U}_{\text{reach}}$ into cell-wise level sets, defined as $\mathbb{S}_{\mathbb{U}_{\text{reach}}}$. 

We next determine a set of proportional values $\rho$, which each encompass information about the coverage of a given `slice' (simplex) of the hull, detailed in Algorithm~\ref{alg:remapping}.
Explicitly, we measure the distance from the boundaries of the full control space $\mathbb{U}$ to each level set $s \in \mathbb{S}_{\mathbb{U}_{\text{reach}}}$, in order to determine how much to augment each component of the hull (lines 4-7). We then use a combination of ray tracing alongside proportional comparison (via $\mathbf{\rho}$) to map the cells in $\mathbb{S}_{\mathbb{U}_{\text{reach}}}$ to their partner cells in $\mathbb{S}_{\mathbb{U}}$ (lines 9-14).

The techniques described up to this point operate within two dimensions only because we constrain the CHO computation to two dimensions for reasons of computational efficiency: the n-dimensional extension of CHO requires that all planar computations performed for the discrete mapping become volumetric computations with a significantly longer runtime. Instead, to extend our approach to interfaces of dimension higher than two, we introduce the \textit{dynamic} part of our ADR approach.  

Specifically, for an $n$-dimensional interface, for each remaining $[n-2]$-dimensions of the signal, we bin the mass of that dimension into $m_z$ levels.
We then run the abovementioned remapping procedure to link $\mathbb{S}_{\mathbb{U}_{\text{reach}}}$ with $\mathbb{S}_{\mathbb{U}}$ for each of the $m_z$ bins, resulting in a \textit{set} of remaps. For the 3-axis
\begin{figure}[H]
\vspace{-3mm}
\begin{algorithm}[H]\captionsetup{labelfont={sc,bf}, labelsep=newline}
\caption{The Discrete Radial Mapping Method \\ 
Given: Set  $\mathbb{C}$ of convex hull simplices, step size $\eta$
}
\label{alg:remapping}
\begin{algorithmic}[1]
\vspace{-0.7mm}
\FOR{$c \in \mathbb{C}$}
    \STATE $\mathbf{c_{\text{mid}}} \equiv \text{ midpoint of } c$ 
     \STATE $\theta \gets \atantwo (c_{\text{mid}, y}, c_{\text{mid}, x})$
     \STATE $\mathbf{r} \gets \mathbf{c_{\text{mid}}}$

    \WHILE{$\mathbf{r} \notin \text{on boundary}$} \STATE $\mathbf{r} \gets \mathbf{r} + ( \theta \cdot \eta ) $
     \ENDWHILE
    \STATE $value \gets 0$
    \FOR{$s \in \mathbb{S}_{\mathbb{U}_{\text{reach}}}$} 
        \STATE $value \gets value ~+~ (1.0-\mathbf{\rho})/\eta$
        \FOR{$cell \in s$} 
            \STATE $cell \gets value$
        \ENDFOR
    \ENDFOR
\ENDFOR
\end{algorithmic}
\end{algorithm}
\vspace{-7mm}
\end{figure}
\noindent joystick example, each remap effectively is conditioned by the extent to which the user is rotating in the third control dimension. As the user teleoperates, the maps are lightweight enough to allow for switching between them in real-time.

\subsection{Deployment Strategies} \label{sec:Deployment Strategies}

Here we propose two deployments of our interface remapping that expands the reachable set of a specific user. $\mathbf{u}^t \in \mathbb{U}$ is a raw interface signal $(u^{t}_{x}, u^{t}_{y}, u^{t}_{z})$ at a given timestep. 

The first deployment, ADR, is the \textit{remapped} controls using our CHO + ADR procedure. The formulation for condition ADR is as follows:
\vspace{-2mm}
\begin{equation}\label{adr} 
    \mathbf{u}^t_{\text{ADR}} = \phi(\mathbf{u}^t). 
\end{equation}

The second deployment, $\text{ADR}_s$, is a \textit{linear combination} of the current $\mathbf{u}^t_{\text{ADR}}$
and that signal smoothed 
over a small moving time window.  $\text{ADR}_s$ is formalized as:
\begin{equation}
    \mathbf{u}^t_{\text{ADR}_s} = (1 - \alpha) \cdot \mathbf{u}^t_{\text{ADR}} + \alpha \cdot \mathbf{\tilde{u}}^t_{\text{ADR}}
\end{equation}

\vspace{-1mm}
\noindent where 
$\mathbf{\tilde{u}}^t_{\text{ADR}}$ is the average remapped control signal over the past $k$ timesteps,
\vspace{-2mm}
\begin{equation}
    \mathbf{\tilde{u}}^t_{\text{ADR}} = \frac{ \sum_{i=t-k}^{t} \mathbf{u}^t_{\text{ADR}}} {k}
\end{equation}
and $\alpha$ is a float value between 0.0 and 1.0. 

We compute $\alpha$ as a function of the \textit{inflection frequency}, $f$, of the user's real-time control signals.
\vspace{-2mm}
\begin{equation}\label{eq:alpha_map}
    \alpha = \begin{cases} 
            0.0, \text{ if } f < f_{lower} \\   
            1.0, \text{ if } f \geq f_{upper} \\
            \text{interpolate}, \text{ o.w.}
            \end{cases} 
\end{equation}
Given $f$, we thus map $\alpha$ proportionally to a pre-determined range of values [$f_{lower}, f_{upper}$].

To compute the inflection frequency $f$, we determine the number of individual traces that occur within a given time window of length $k$ sampled at $f_{\text{rate}} \text{ Hz}$. We define a \textit{trace} a trajectory segment with a change in angle of $\geq 30 \degree$ from the preceding point in the trajectory. When a trace makes a sudden deviation---corresponding to a high contemporaneous jerk in the signal---we clip that trace and count it as an inflection before tracking the next trace. 
Additionally, $f$ is used as the \textit{omission} criteria for determining the reachable set $\mathbb{U}_{\text{reach}}$. Points in the user distribution where $f > 8.0$ are considered to be \textquote{unstable} and not included in $\mathbb{U}_{\text{reach}}$. 

\section{EXPERIMENTAL DESIGN}\label{sec:exp_design}

Here we detail our experimental setup and protocol.

\subsection{Hardware and Materials}
We used an APEM Inc. IPD Ultima
3-axis Joystick for this study. Signals were logged at a frequency of $60 \text{ Hz}$. For CHO, we used \textit{Qhull}~\cite{barber1996quickhull} in python. For the virtual 3D-center-out task environment, we used \textit{pyglet}.

\subsection{Experimental Parameters}

For this study, we chose the resolution parameters $\eta=0.5$, $m_x = m_y = 160$, and $m_z = 5$.
We discretized the z-axis into 5 layers ($m_z$), to optimize 
the tradeoff between fidelity and computation time. These parameters kept computation time under 25 minutes. 
Parameters controlling the inflection sampling frequency were chosen as $f_{lower} = 8.0 \text{ Hz, } f_{upper} = 20.0 \text{ Hz, } f_{\text{rate}} = 40.0 \text{ Hz, and } k = 20$  based on the literature surrounding the frequency of tremor~\cite{vernooij2015complete}.

\subsection{Participants} \label{sec:Participants}
Eight participants were recruited: 5 stroke survivors (55.8 $\pm$ 10.2 years of age)
and 3 individuals with Spinal Cord Injury (SCI) (41.0 $\pm$ 20.5 years of age).
All participants were screened for ability to operate a 3-axis joystick (twisting), and presence of motor impairment in their dominant arm. All participants used their dominant arm. One stroke participant was not able to complete the tasks due to severely reduced mobility in their dominant arm, and another stroke participant was unable to complete the tasks due to fatigue from teleoperation leading to extended rest intervals. Data from these two participants is not included in the analysis; we present data from 6 participants (3 stroke, 3 SCI). All participants gave their informed, signed consent to participate in the experiment which was approved by Northwestern University’s Institutional Review Board.

\subsection{The 3D-Center-Out Reaching Task}

\begin{wrapfigure}[7]{r}{2.8cm}
\vspace{-1.2cm}
    \includegraphics[width=1\hsize]{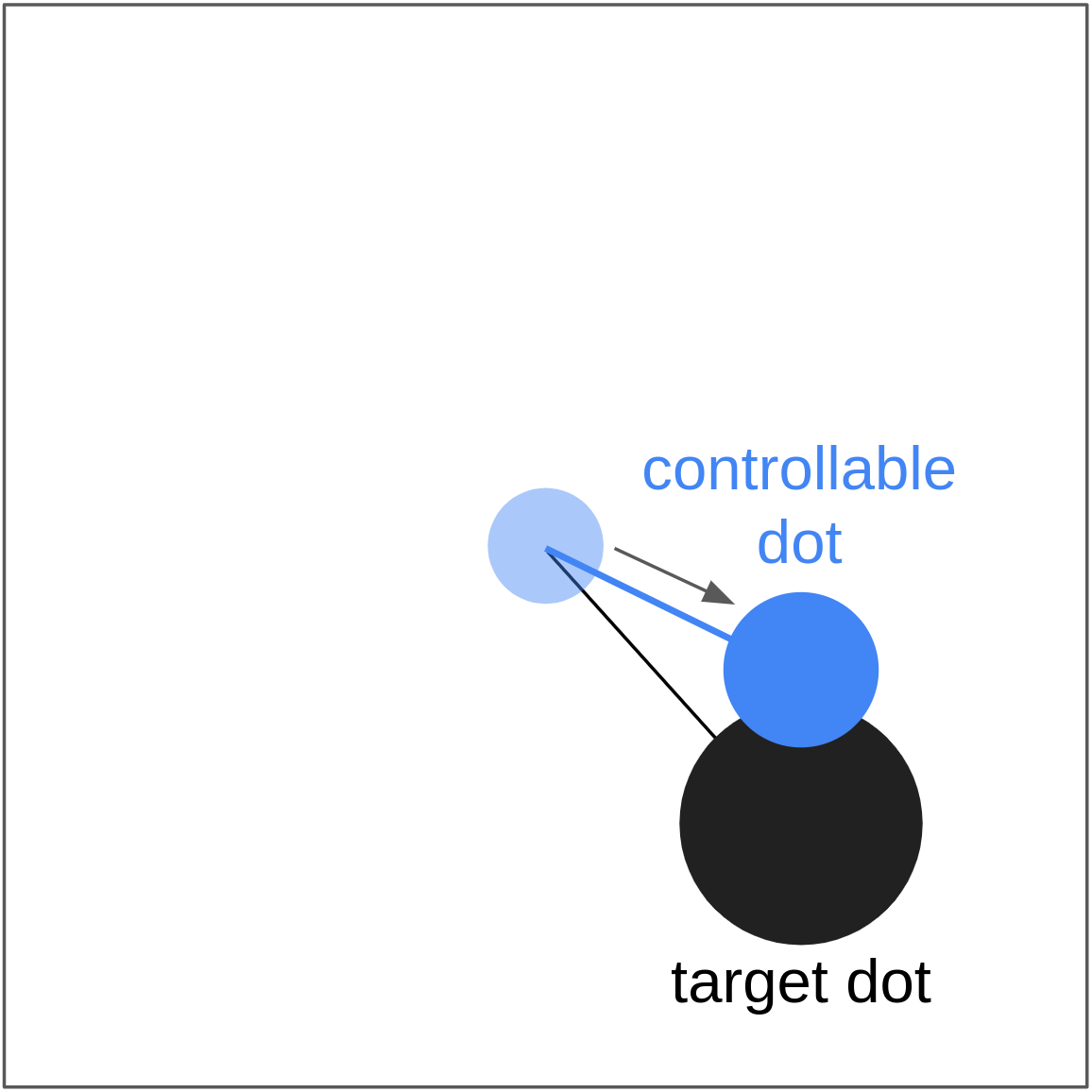}
    \caption{
    3D-center-out reaching task \vspace{1mm}}
    \label{fig:painting_task}
\end{wrapfigure}
To collect data concerning the reachability of the control space for a given user during teleoperation, we designed a virtual center-out reaching task in 3D
which consisted of a series of target dots (2D position + 1D size) and one controllable dot (Fig.~\ref{fig:painting_task}). 

Each of the target dots were presented in random order, 
and parameterized by ($x$, $y$, $z$), where $z$ corresponds to the dot's size. The user was able to control a separate dot (`the controllable dot') with the 3-axis joystick; specifically, the user was able to deflect the joystick to cause the controllable dot to move translationally and they were also able to rotate the joystick to change the dot's size. The participant goal was first to teleoperate the controllable dot until it was in the same location as and the same size as the current target dot, and then to hold the controllable dot in that position for 2 seconds (as a measure of stability).

\subsection{Experimental Procedure}

Study sessions started with a training phase, where participants were given a 3D-center-out task with 25 targets randomly sampled from the set of all possible targets. Participants were familiarized with how the joystick control maps to the size and position of the controllable dot, as well as the task's completion conditions. After this, participants first completed 
the 3D-center-out task (125 targets) without any remapping (condition $\neg$ADR).

The resulting distribution of participant control commands were used to compute their reachable space and build their customized three-dimensional map $\phi$. The generation of the three-dimensional map can be a lengthy process depending on the discretized resolution of $\mathbb{U}$ and the number of binned hulls 
along the third dimension. Hence, the choice of a 160 $\times$ 160 discrete grid as well as 5 separate hulls was determined experimentally to keep computation time under 20 minutes. 

After the remap was computed, participants performed two additional rounds of the full 3D-center-out task under both ADR and $\text{ADR}_s$, 
with a random presentation balanced across participants. Every 25 targets, participants had the option to take a break to combat fatigue.
As stated in 
Section~\ref{sec:Participants}, rates of fatigue during teleoperation were variable between participants, and the optional rest periods included in the 3D-center-out task were utilized to varying degrees across participants. As such, participants completed this protocol over 2 or 3 sessions, depending on their rates of progress.

\subsection{Task Completion Metrics and Evaluation} \label{sec:Task Completion Metrics}  
Our bias-aware teleoperation remapping was evaluated through the computation of a number of metrics
on both raw and remapped user commands. Comparisons were made between remapped controls (ADR and $\text{ADR}_s$) and unmapped controls ($\neg$ADR), as well as between raw and remapped controls under the ADR and $\text{ADR}_s$ conditions.
\begin{itemize}
    \item \textit{Change in target completion}:
    A per-target pairwise comparison of the number of targets completed across remapping conditions.
    \item \textit{Change in path efficiency}:
    The difference 
    in path length between
    trials with remapping (ADR or $\text{ADR}_s$) as compared to  trials with no remapping ($\neg$ADR).
    \item \textit{Time to first reach target}: Time taken for the user controlled dot to first fall within threshold of the target.
\end{itemize}

\begin{figure*}
    \vspace{1.2mm}
    \centering
        \includegraphics[width=0.99\textwidth]{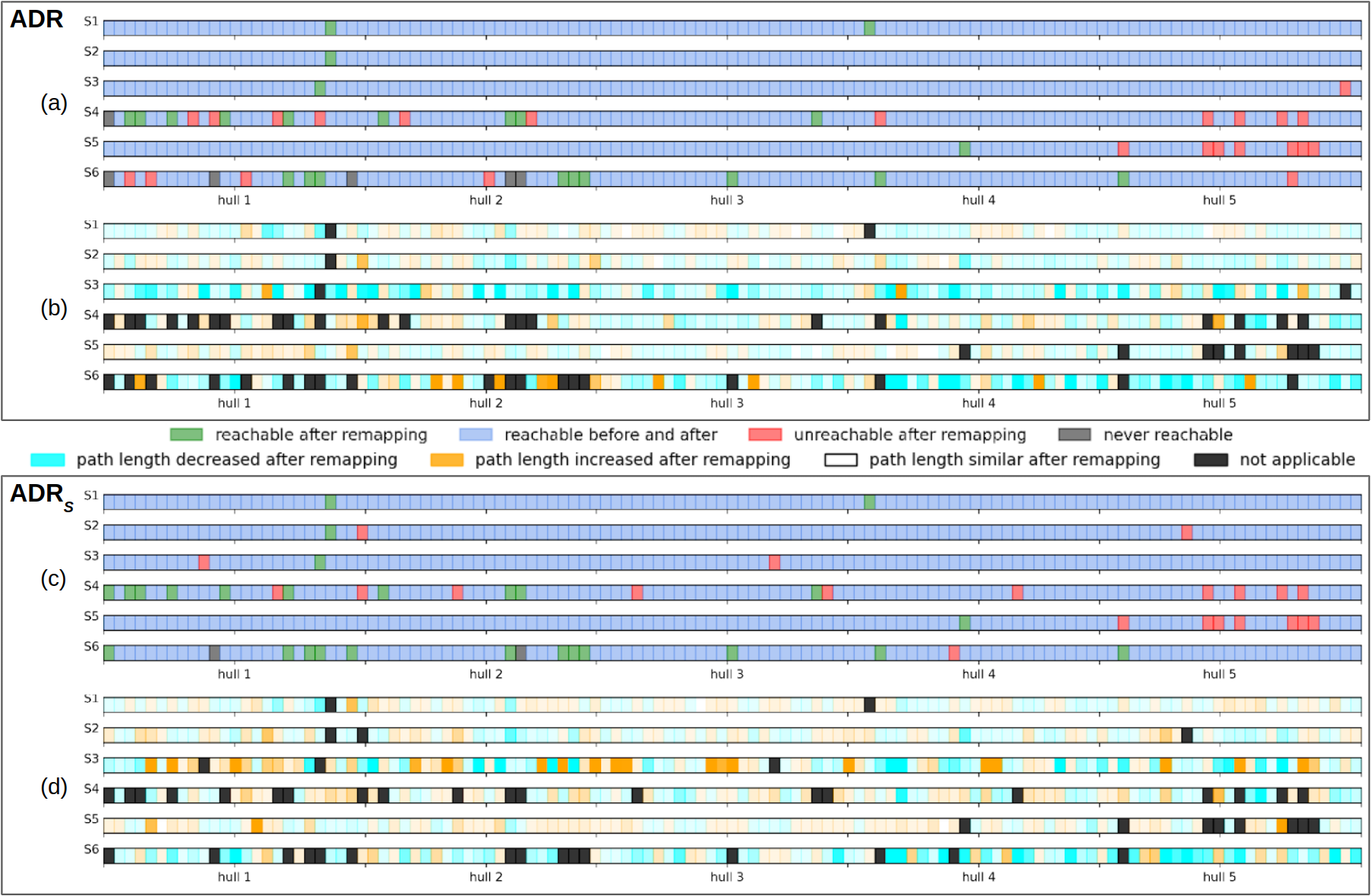}
        \vspace{0.5mm}
        \caption{Overview of effect of remapping conditions on task performance.
        Effect of \textbf{(a)} ADR  and \textbf{(c)} $\text{ADR}_s$  on target completion. Effect of \textbf{(b)} ADR  and \textbf{(d)} $\text{ADR}_s$ on path efficiency for targets that could be reached under both $\neg$ADR and, respectively, ADR or $\text{ADR}_s$. (Targets not reached under both conditions are represented in black.) Higher intensity of colors maps to greater change---increase (orange) or decrease (blue)---in path length.}
        \label{fig:AC_effects}
        \vspace{-6mm}
\end{figure*}

\section{RESULTS}\label{sec:results_and_discussion}

We begin
by presenting an overview of the effect of our remapping conditions on each participant (Fig.~\ref{fig:AC_effects}).

\subsection{Reachability}

In Fig.~\ref{fig:AC_effects}, for both remapping conditions ADR and $\text{ADR}_s$ each target is presented as a colored square, where the color represents the effect that remapping had on the target's reachability by the participant (panels (a) and (c)). Green indicates a location reachable \textit{only after} remapping; blue a location reachable \textit{before and after} remapping; red a location \textit{previously reachable} and then unreachable after remapping; and gray a location \textit{unreachable} both before and after remapping.

We break down these reachability results per participant. For S6, both ADR and $\text{ADR}_s$ had a net positive effect on their target completion, with $\text{ADR}_s$ being particularly effective in helping them reach more targets ($\sim 10\%$ increase).  For S4, both ADR and $\text{ADR}_s$ improved reachability of $\sim 8\%$ of the targets, but also hindered their ability to reach a similar number of targets that they were previously able to reach. This is seen more in the first 2 hulls, and likely reflects an asymmetry in S4's actuation of the joystick in the twist axis, with a bias towards one direction of twist over the other. 

Participants S1-3 were able to successfully complete $>98\%$ of targets before any remapping occurred. Under ADR and $\text{ADR}_s$, no significant improvement in target completion was observed for these participants. Participant S5 was able to successfully complete $>98\%$ of targets before remapping. However $6\%$ of targets become unreachable after remapping, for both of ADR and $\text{ADR}_s$. Based on observations during their trial, the criteria for omission $f$ (\ref{sec:Deployment Strategies}) may have inaccurately omitted points from the reachable space in computing the remap, resulting in a larger than desired change to their control map under ADR.

\subsection{Efficiency}
We take a deeper look at targets participants were already able to reach under $\neg$ADR and whose reachability was not impacted by the remapping. For these targets, we present the change in path length traversed
after remapping (panels (b) 
\begin{figure}[H]
\vspace{-2.2mm}
\begin{table}[H]
    \caption{Tally of each target completion condition. Column headings correspond with colors in  Fig.~\ref{fig:AC_effects}.}
    \label{tab:tally_completion}
    \begin{center}
        \begin{tabular}{c c c c c c}
        \hline
        Participant ID & Condition & Green & Blue & Red & Gray \\
        \hline
        \multirow{2}{*}{S1} & ADR & 2 & 117 & 0 & 0 \\
                            & $\text{ADR}_s$ & 2 & 117 & 0 & 0 \\
        \hline
        \multirow{2}{*}{S2} & ADR & 1 & 118 & 0 & 0 \\
                            & $\text{ADR}_s$ & 1 & 116 & 2 & 0 \\
        \hline
        \multirow{2}{*}{S3} & ADR & 1 & 117 & 1 & 0 \\
                            & $\text{ADR}_s$ & 1 & 116 & 2 & 0 \\
        \hline
        \multirow{2}{*}{S4} & ADR & 9 & 98 & 11 & 1 \\
                            & $\text{ADR}_s$ & 10 & 99 & 10 & 0 \\
        \hline
        \multirow{2}{*}{S5} & ADR & 1 & 111 & 7 & 0 \\
                            & $\text{ADR}_s$ & 1 & 111 & 7 & 0 \\
        \hline
        \multirow{2}{*}{S6} & ADR & 9 & 100 & 5 & 5 \\
                            & $\text{ADR}_s$ & 12 & 104 & 1 & 2 \\
        \hline
        \end{tabular}
    
    \end{center}
\end{table} 
\vspace{-10mm} 
\end{figure}
\noindent and (d)). The effect of remapping on the path length of target trajectories provides some insight about the total effort
required from a participant, on average, to complete the 3D-center-out task. 
For 4 out of 6 participants, we saw that path length was shorter under remapping conditions in extreme hulls---that is, hulls 
which correspond to larger simultaneous deflection and twisting motions of the joystick.

\begin{figure}
    \centering
    \vspace{2.5mm}
    \includegraphics[width=0.95\linewidth]{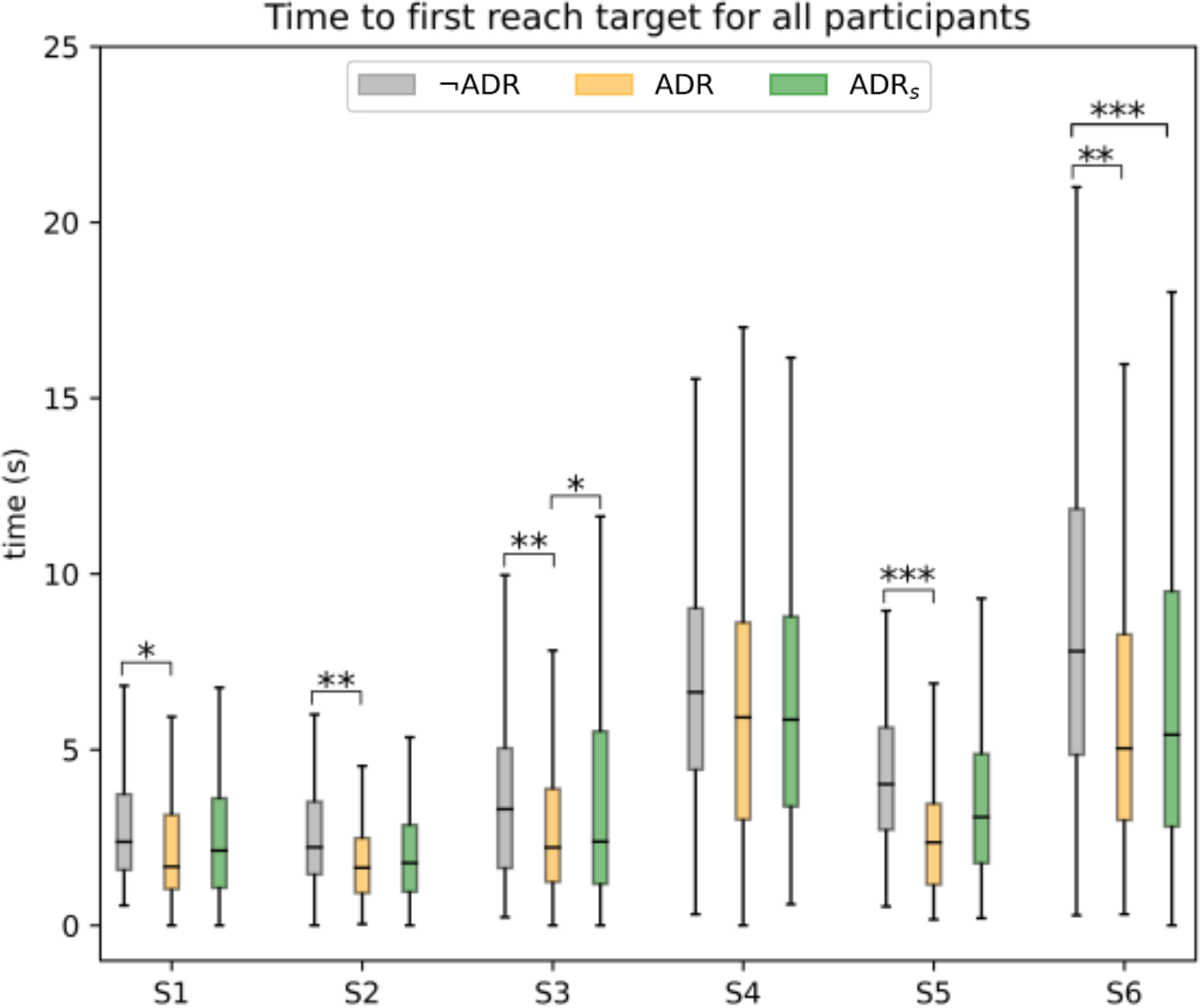}
    \caption{Time to first reach targets across remapping conditions and participants. Statistical significance was computed using a $t$-test.}
    \vspace{-6mm}
    \label{fig:time_first_reach}
    \vspace{-0.7mm}
\end{figure}

We also present in  Fig.~\ref{fig:time_first_reach} the
time taken for each participant to first reach a target, averaged over all targets. Note that the high standard deviation is not unexpected, because the distance between the starting position of each trial (neutral joystick position) and the respective ending positions differ between targets. 
Furthermore, as each participant saw 
 the same targets, differences between them 
are due to inter-
\noindent participant variability in response characteristics.

We found that the time taken to first reach the targets 
decreased significantly under ADR for 5 out of 6 participants (S1-3, S5-6), indicating a positive effect of our remap. This is likely due to the fact that, under ADR, a lesser magnitude of interface actuation is required by the user to achieve a similar reachability of the control space as in $\neg$ADR. 

A similar decrease was not observed for trials under $\text{ADR}_s$. 
This might have been due to the standardized sampling of $f$ and its effect on the resultant computation of $\alpha$; as optimal frequency thresholds may differ for participants, its effect of control blending may have been diluted.
Another potential reason 
was the use of a moving average to compute the resultant control signals, which introduces a response lag to user input. 

\section{DISCUSSION \& FUTURE WORK} \label{sec:discussion}

From this pilot study, across participants, we found an improvement in interface actuation efficiency under remapping ADR, and a range improvement in reachability with the addition of either remapping condition.

These results suggest that further tuning parameters within our framework---resolution of the remap, omission frequency $f$, rate $f_{rate}$, step size $\eta$---will be most effective customized for each user. Such individualized tuning 
will likely require a more involved data collection phase for building the bias profile. 
The optimization could be realized by embedding our remapping procedure within a machine learning framework.

When a given participant had an almost fully reachable set to begin with, there were cases of remapping hindering their target achievement. This could be because such participants already possess a complete map of the full control space, and (small) adjustments to that map via our remapping procedure often showed up as subtle discontinuities during operation.

It is important to note that the procedure outlined in this study will not account for potential `holes' in a user's reachable space, nor would it be able to generate more complex nonlinear remappings. We did not observe these during our study, though it is conceivable that some individuals may have unreachable regions 
embedded within the space that our method would deem as the reachable set $\mathbb{U}_{\text{reach}}$; 
due to our use of CHO to determine the reachable set, which requires convexity. One possible approach to combat this requirement would be to partition the control space into multiple locally convex regions,  allowing for more complex and discontinuous reachable sets to be parameterized. In future work, we hope to extend our remapping model to more flexibly handle high-dimensional interfaces, nonlinear mappings (e.g.,~orientation mismatch), as well as to incorporate user-specific parameter tuning for our remapping procedure.

\section{CONCLUSIONS}
In this study, we have presented a novel method for efficiently computing a customized map between a user's available space of 
interface actuation and the full control space of the interface in question. The utility of our remapping was evaluated in a study with human participants. We found that the total control actuation required to reach the same space post-remapping was generally reduced, suggesting a beneficial effect of our remapping procedure, and our results show improvements in task completion efficiency.

\addtolength{\textheight}{-12cm}   





\section*{ACKNOWLEDGMENTS}
We gratefully acknowledge the support of this work by the U.S. Office of Naval Research under the Award Number N00014-16-1-2247.



\bibliography{IEEEabrv,ref/ICORR2022_BiasMetricEvaluation.bib}

\end{document}